\documentclass[a4paper,12pt]{article}
\usepackage[dvips]{graphicx}
\usepackage{eufrak}

\title{On Sommerfeld precursor in a Lorentz medium}
\author{Adam Ciarkowski \\ \textit{\small
Institute of Fundamental Technological Research}\\
\textit{\small Polish Academy of Sciences}}

\def\o{\omega}
\def\d{\delta}
\def\t{\theta}

\def\bet{\mathcal{B}}
\def\b{\beta}
\def\l{\lambda}
\def\z{\zeta}

\begin{document}

\date{}

\maketitle

\begin{abstract}
A one-dimensional electromagnetic problem of Sommerfeld precursor
evolution, resulting from a finite rise-time signal excitation in
a dispersive Lorentz medium is considered. The effect of the
initial signal rate of growth as well as of the medium dumping on
the precursor shape and its magnitude is discussed. The analysis
applied is based on an approach employing uniform asymptotic
expansions. In addition, new approximate formulas are given for
the location of the distant saddle points which affect local
frequency and dumping of the precursor. The results obtained are
illustrated numerically and compared with the results known from
the literature.
\end{abstract}

\section{Introduction}

Fundamental investigations on EM signal propagation in the Lorentz
model of a dispersive medium are due to Sommerfeld \cite{so;14}
and Brillouin \cite{br;14,br;60}. These authors revealed that in
addition to the main signal propagating in the medium, two
precursors are formed which precede the signal. The front of the
fastest (Sommerfeld) precursor propagates in the medium with the
velocity of light. The instantaneous oscillation frequencies of
the precursors and their local dumping are directly related to the
locations in the complex frequency plane of the corresponding
saddle points in the integral representation of the signal. Those
locations vary with space and time, and are governed by the saddle
point equation (SPE) in Eq.~(\ref{e7}), requiring that the phase
function in the integrand in that representation be stationary.
Analysis of the equation shows that there are two pairs of
dominant saddle points, the distant and the near ones, responsible
for the first (referred to as Sommerfeld) and the second
(Brillouin) precursors, respectively. The saddle points in each
pair are located symmetrically with respect to the frequency
imaginary axis, the fact being related to the causality principle.

Since SPE seems to be not solvable in a closed form, attempts have
been made to solve it in an approximate manner. Brillouin's
approach \cite{br;60} consisted in replacing the complex index of
refraction in SPE by its expansion in powers of frequency, and
then solving the simplified equation for frequency. As a result, a
simple, approximate formula was found relating complex frequency
to the distance and time coordinates. In the case of the
Sommerfeld precursor, the applicability of this map was confined
to the vicinity of the front of the precursor. Unfortunately,
Brillouin, having at his disposal only a non-uniform asymptotic
method, could not effectively describe the front evolution, as it
corresponds to coalescence of two distant saddle points at
infinity, the case not treatable with the method he used. This
deficiency was removed by Bleistein and Handelsman \cite{hb;69}
who developed a uniform asymptotic approach that extended the
validity of asymptotic considerations to the precursor front.

Recently, the problem of asymptotic analysis of signal propagation
in dispersive media has been reexamined by many authors. Extensive
research in this field is due to Oughstun and Sherman \cite{ou;97}
and Kelbert and Sazonov \cite{ks;96}. In a recent work,
\cite{dz;98} Dvorak, Ziolkowski and Felsen offered a new hybrid
approach combining both the asymptotic and FFT methods, the former
being responsible for extreme parts of the signal frequency
spectrum.

In this work we reconsider the asymptotic model of signal
propagation in a Lorentz medium, and concentrate on the Sommerfeld
precursor. We assume that the Sommerfeld precursor is excited in
the Lorentz medium by a sine modulated signal, with its envelope
described by a hyperbolic tangent function. Such a representation
provides a convenient model for signals with finite rise time.
Unlike the Oughstun-Sherman study ((\cite{ou;97}, Secs.~4.3.4,
7.2.7 and 7.3.7), where the signal envelope was described by an
everywhere non-zero smooth function, our initial signal has the
form of an abruptly switched modulated sine signal, i.e.\ it
vanishes identically for $t<0$ and its envelope is non-zero for
$t>0$. At $t=0$ the derivative of the envelope suffers a step
discontinuity. We study the influence of both the medium and the
initial signal characteristics, including medium damping $\d$ and
signal speed factor $\b$, on the evolution of the Sommerfeld
precursor in the medium. In particular, we analyze how the speed
factor affects the shape and the magnitude of the precursor
excited by both the slow and the fast growing incident signals. We
also obtain a simple approximation for the precursor damping
factor. Finally, we provide a new approximation to the location of
the saddle points, which is more accurate that those known in the
literature.

The results obtained here may appear to be useful e.g.\ in designs
employing fast Sommerfeld precursors as signals triggering the
electronic devices designed to process the main signal.

\section{The propagation problem and its exact solution}

We consider the 1D problem of EM signal propagation in a Lorentz
medium characterized by the complex index of refraction
\begin{equation}\label{e1}
  n(\o) = \left( 1 - \frac{b^2}{\o^2 - \o_0^2 + 2 i
  \delta \o}\right)^{1/2},
\end{equation}
where $b^2=4\pi N e^2/m$ is the plasma frequency of the
medium, $N$, $e$ and $m$ are the number of electrons per
unit volume, electron charge and its mass, respectively,
$\delta$ is a damping constant and $\o_0$ is the resonant
frequency.

In the plane $z=0$ of the Cartesian coordinate system $\{x,y,z\}$,
the field component $E_x(0,t)$, henceforth denoted by $E_0(t)$, is
assumed to be given. It is described by a function, which has the
form of a finite rise-time, sine modulated signal
\begin{equation}\label{e2}
   E_0(t)=\left\{
     \begin{array}{ll}
        0 &               t<0 \\
        \tanh(\b t) \sin(\o_ct) & t\ge 0.
     \end{array}
    \right.
\end{equation}
The large positive coefficient $\b$ determines how rapidly the
signal turns on, and $\o_c$ is a fixed carrier frequency. It is
also assumed that no EM sources are present at
$z\rightarrow\infty$. In \cite{ou;97} a different signal was
considered, also employing a tangent hyperbolic function, which,
unlike (\ref{e2}), did not vanish for times $t<0$.

In general, the problem consists in finding the field in the
half-space $z>0$ and for time $t>0$. The solution to this mixed,
initial-boundary value problem for the Maxwell equations takes the
form \cite{ac;97}
\begin{equation}\label{e3}
   E(z,t)=\frac{1}{2\pi}\int_{i a-\infty}^
   {i a+\infty}g(\o\;;\b,\o_c)
   \exp{\left[i\frac{z}{c}\Psi(\o,\t)\right]}\,d\o,
   \nonumber
\end{equation}
where the constant $a$ is greater than the abscissa of absolute
convergence for $E_0(t)$. The amplitude and phase functions
$g(\o\;;\b,\o_c)$ and $\Psi(\o,\t)$, respectively, are given by
\begin{eqnarray}\label{e3a}
\lefteqn{g(\o\;;\b,\o_c)=} \nonumber \\[1ex] & &
\frac{1}{2}\left\{\frac{i}{\b}\bet\left[-\frac{i(\o-\o_c)}{2\b}\right]
+\frac{1}{\o-\o_c}
-\frac{i}{\b}\bet\left[-\frac{i(\o+\o_c)}{2\b}\right]
-\frac{1}{\o+\o_c}\right\}
\end{eqnarray}
and
\begin{equation}\label{e4}
   \Psi(\o,\t)=\o[n(\o)-\t].
\end{equation}
The beta function is defined through the psi function as
\begin{equation}\label{e5}
  \bet(s) = \frac{1}{2}\left[\psi\left(\frac{s+1}{2}\right) -
  \psi\left(\frac{s}{2}\right)\right].
\end{equation}
(For the definition and properties of the psi function see
\cite{as;64}, Sec.~6.3.) The beta function is related to the
envelope of $E_0(t)$ via the Fourier transformation
\begin{equation}\label{e5a}
\int_0^\infty \tanh{\b t} e^{i\o t}\, dt =
\frac{1}{\b}\bet\left(-\frac{i\o}{2\b}\right) - \frac{i}{\o}.
\end{equation}
Finally,
\begin{equation}\label{e6}
   \t=\frac{c t}{z}
\end{equation}
is a dimensionless parameter that characterizes a
space-time point $(z,\,t)$.

In this work we confine our interest to one component of
the general solution -- the Sommerfeld precursor. A
suitable approach to extract this partial field and study
its dynamics is to evaluate the integral (\ref{e3})
asymptotically as $z\rightarrow\infty$. At the precursor
onset, i.e.\ as $\t\rightarrow 1^+$, the distant saddle
points in the complex $\o$ plane meet at infinity to form
a saddle point of infinite order. The valid asymptotic
procedure that handles this case is a special instance of
application of the general asymptotic theory developed by
Bleistein and Handelsman and designed to uniformly
evaluate integrals with nearby critical points
\cite{bh;75}. (It can also be used in case of more than two
coalescing critical points. The case of three critical
points: a pole, a branch point and a saddle point was
studied in \cite{ac;89}.)

Dynamics of the saddle points, essential in asymptotic
considerations, is governed by the saddle point equation (SPE)
\begin{equation}\label{e7}
   n(\o)+\o n^{\prime}(\o)-\t=0.
\end{equation}
This equation results from the requirement that the phase
(\ref{e4}) should be stationary. It has the form $\t=f(\o)$. What
we need is the inverse function $\o=f^{-1}(\t)$. Approximate
solutions to the latter equation were obtained by Brillouin
(\cite{br;60}), Kelbert and Sazonov (\cite{ks;96}) and Oughstun
and Sherman (\cite{ou;97}). In Sec.~\ref{s} we present another
approximate solution, which is more accurate than the solutions
known in the literature.

\section{Uniform asymptotic representation for the
Sommerfeld precursor}

The phase function $\Psi(\o,\t)$ has the saddle point of infinite
order at infinity (see \cite{bh;75}, Ch.\ 9). In this case the
classical asymptotic methods break down, which implies that they
cannot be used to describe the precursor behavior at its front
(i.e.\ for $\t\rightarrow 1^+$). Fortunately, the uniform
approach, as proposed in \cite{bh;75}, can be effectively used.
The term "uniform" means that the resulting asymptotic expansion
is valid for any locations of the far saddle points in the $\o$
complex plane, including the case where the points (symmetrical
with respect to the imaginary axis) coalesce at infinity to create
one saddle point of infinite order. As a consequence, the
expansion holds for any $\t\ge 1$, in particular at the precursor
front. By using the methods described in \cite{bh;75}, Sec.\ 9.5,
we arrive at the following uniform asymptotic
representation\footnote{This representation is equivalent, but
simplified in form, to that presented in \cite{ac;00}. In
\cite{ac;00} the factor 1/4 in Eq.(4.5) should be replaced by
1/8.} for the Sommerfeld precursor as $z\rightarrow\infty$:
\begin{eqnarray}\label{e24}
\lefteqn{E^S(z,t)\sim -\exp{\{-\l\; \hbox{Im}[\Psi(\o^+,\t)]\}}}
\hspace{3cm}  \\[1ex] & &{}\times\{\hbox{Re}[G(\o^+; \b,\o_c)]
J_1[-\l\;\hbox{Re}(\Psi(\o^+,\t))] \nonumber \\[1ex] & &
{}+\hbox{Im}[G(\o^+; \b,\o_c)]
J_2[-\l\;\hbox{Re}(\Psi(\o^+,\t))]\}, \nonumber
\end{eqnarray}
where
\begin{equation}\label{e25}
G(\o^+;\b,\o_c)=\sqrt{\frac{\hbox{Re}[\Psi(\o^+,\t)]}
{\Psi_{\o\o}(\o^+,\t)}}\;g(\o^+;\b,\o_c),
\end{equation}
$J_1(\cdot)$ and $J_2(\cdot)$ are Bessel functions of the order
$1$ and $2$, respectively, and $\l=z/c$.

In this paper we do not consider possible transition from the
precursor to the main signal, which may happen for very high
carrier frequency $\o_c$. This occurs when the poles
$\o=\pm\o_c-2ik\b$, $k=0,1,2,\ldots$ of the beta function are
crossed during deformation of the original contour of integration
in (\ref{e3}) to the SDP through $\o=\o^\pm$.

A fundamental question that now arises is how the speed parameter
$\b$ in (\ref{e2}) and the damping parameter $\d$ affect the
precursor dynamics in the medium. Below we try to answer this
question.

\subsection{Dependence of the precursor on $\b$}
Let the Lorentz medium considered be described by Brillouin's
parameters
\begin{equation}\label{e26}
b=\sqrt{20.0}\times 10^{16} s^{-1}, \quad \o_0=4.0\times
10^{16} s^{-1}, \quad \d=0.28\times 10^{16} s^{-1}
\end{equation}
and let us additionally choose
\begin{equation}\label{e27}
\o_c=2.0\times 10^{16} s^{-1}, \quad \l=5.0\times 10^{-15} s.
\end{equation}
Assume further that $\b=1.0\times 10^{14} s^{-1}$. The dynamic
behavior of the Sommerfeld precursor field corresponding to this
data, is shown in Fig.\ 1.

\begin{figure}
   \centering
      \includegraphics[width=.7\textwidth]{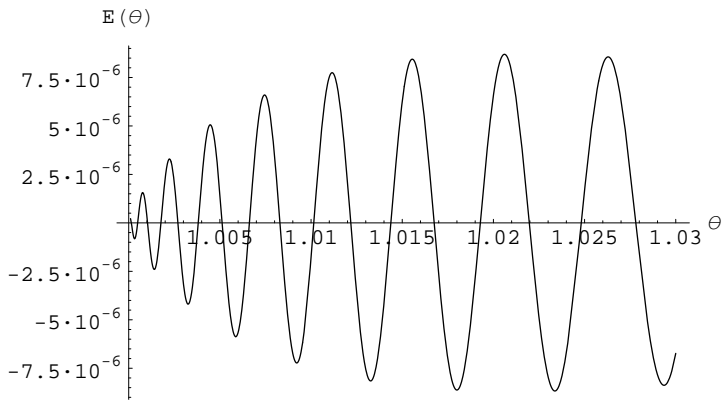}
      \caption{\textit{Dynamic behavior of the
       Sommerfeld precursor in the Lorentz medium described by
       Brillouin's parameters. $\o_c=2.0\times 10^{16} s^{-1}$,
       $\l=5.0\times 10^{-15} s^{-1}$ and $\b=1.0\times 10^{14}s^{-1}$.}}
\end{figure}

\vspace{1ex} \noindent If now $\b$ is increased, by one order or
two, it is can be shown from (\ref{e24}) that the oscillations in
Fig.\ 1 are increased by the same factor, while the shape of the
precursor is preserved. If, however, $\b$ takes much higher
values, such as $\b=1.0\times 10^{19} s^{-1}$ or more, the
precursor shape distinctly changes (see Fig.~2), and the
oscillation amplitudes virtually remain at the same, relatively
high level as $\b$ further increases.

\begin{figure}
   \centering
      \includegraphics[width=.7\textwidth]{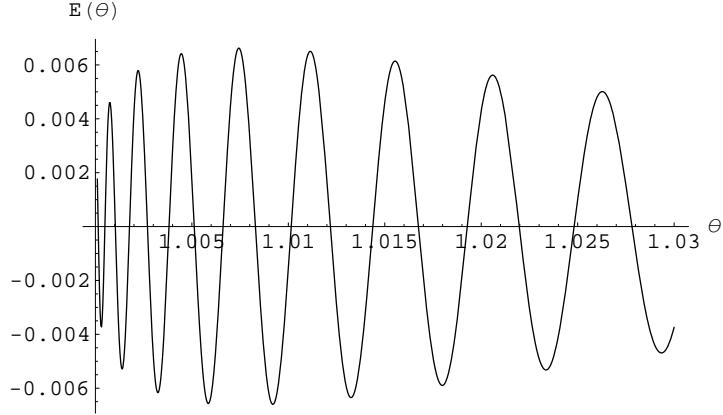}
      \caption{\textit{Dynamic behavior of the
       Sommerfeld precursor in the Lorentz medium described by
       Brillouin's parameters. $\o_c=2.0\times 10^{16} s^{-1}$,
       $\l=5.0\times 10^{-15} s^{-1}$ and $\b=1.0\times 10^{19}s^{-1}$.}}
\end{figure}

This interesting behavior can be explained by studying properties
of the function $G(\o^+;\b,\o_c)$. The plot of $G(\o^+;\b,\o_c)$
as a function of $\b$ with fixed remaining arguments is presented
in Fig.\ 3.

\begin{figure}
   \centering
      \includegraphics[width=.7\textwidth]{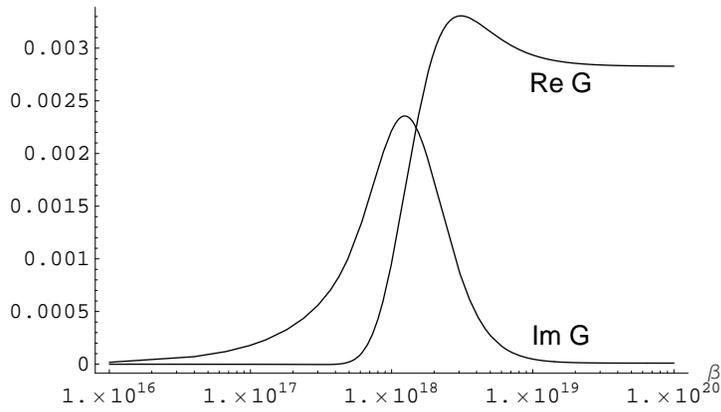}
      \caption{\textit{Dependence of the real (solid
       line) and imaginary (dashed line) parts of the function
       $G(\o^+;\b,\o_c)$ on the parameter $\b$. Calculated for
       $\t=1.0001$ and $\o_c=2.0\times 10^{16} s^{-1}$.}}
\end{figure}

\vspace{1.5ex} The value $1.0001$ of the variable $\t$ was chosen
what corresponds to the close vicinity of the precursor front
(cf.\ Fig.\ 1 and Fig.\ 2). It is seen from the plot that there
are three characteristic regions of $G(\o^+;\b,\o_c)$ variation.
If for fixed $\o^+$ the parameter $\b$ is relatively small then
the real part of function $G(\o^+;\b,\o_c)$ is virtually zero and
the essential contribution to the precursor is due to its
imaginary part. This contribution increases in value with rising
$\b$ until about $\b=4.0\times 10^{17} s^{-1}$. This is the first
region in which the precursor oscillation is described by the
Bessel function $J_2$. In the second, transitory region, the real
part of $G$ grows rapidly, and at about $\b=2.0\times 10^{19}
s^{-1}$ it settles down at a virtually constant level. The
imaginary part reaches its maximum and then steadily decreases to
zero. Here the contribution of the Bessel function $J_1$ takes
over. Finally, in the third region the real part of $G$ remains
nearly unchanged and the imaginary part vanishes. Now the
contribution of $J_1$ dominates and that of $J_2$ is to be
neglected. One can verify that with increasing $\t$, the third
region broadens, thus pushing the second region in the direction
of smaller $\b$.

Below we analyze the precursor behavior in the first and the third
regions in more detail. To make the results as simple as possible,
we shall employ Brillouin's approximation
\begin{equation}\label{e28}
\o^+(\t)\approx \frac{b}{\sqrt{2(\t-1)}}-2 i \d.
\end{equation}
The use of this approximation is justified in the vicinity of the
precursor front.

\vspace{1ex} \textsf{The case of relatively small $\b$}

\vspace{1ex} \noindent Here, by choosing sufficiently small $\t$
both arguments in functions $\bet$ in (\ref{e3a}) can be made
arbitrarily large. In this case the function $G$ in (\ref{e25})
may be simplified by substituting $\bet$ for its asymptotic
expression
\begin{equation}\label{e29}
\bet(u)=\frac{1}{2u}+\frac{1}{4u^2}+O(u^{-4}), \qquad u\rightarrow
\pm i\infty,
\end{equation}
valid also in some sectors centered around the rays $\arg{u}=\pm
i\pi/2$. Then,
\begin{equation}\label{e30}
G(\o^+;\b,\o_c)\sim \sqrt{\frac{\hbox{Re}(\Psi(\o^+,\t))}
{\Psi_{\o\o}(\o^+,\t)}}\cdot\frac{-4i\b
\o_c}{(\o^{+^2}-\o_c^2)^2}.
\end{equation}
If we expand this in fractional powers of $\t-1$ and retain the
leading terms, we arrive at
\begin{equation}\label{e31}
G(\o^+;\b,\o_c)\sim
\frac{48\sqrt{2}\b\d\o_c(\t-1)^{3/2}}{b^3}-
\frac{8i\b\o_c(\t-1)}{b^2}.
\end{equation}
Since $\d\sqrt{\t-1}/b$ is a small quantity, we can safely retain
only the imaginary term. We see that $G(\o^+;\b,\o_c)$ is
proportional to $\b$. This fact confirms the observed behavior of
$G$ in the first region shown in Fig.\ 3.

Let us now consider the envelope of the Sommerfeld
precursor. It is obtained by replacing the Bessel
functions in (\ref{e24}) by their envelopes. A good
approximation for these envelopes is $\sqrt{2/(\pi x)}$,
where $x$ denotes the argument of a Bessel function. Thus,
by (\ref{e24}), the envelope of the precursor can be
approximated by
\begin{eqnarray}\label{e32}
\lefteqn{\widetilde{E}^S(z,t)\approx -\exp\{{-\l
\hbox{ Im}[\Psi(\o^+,\t)]\}} } \nonumber \\
& & {} \times
\sqrt{\frac{-2}{\pi\l\;\hbox{Re}[\Psi(\o^+,\t)]
}}\{\hbox{Re}[G(\o^+; \b,\o_c)] + \hbox{Im}[G(\o^+;
\b,\o_c)] \}.
\end{eqnarray}
Naturally, the notion of the envelope applies here to $\t$ greater
than the first extreme of $J_2(x)$, occurring at $x\approx 3.054$.
In our case it corresponds to $\t\approx 1.00009$.

If we are interested in the precursor front only, we can use
(\ref{e28}) in (\ref{e32}), replace $G(\o^+;\b,\o_c)$ by RHS of
(\ref{e31}), and expand the result in fractional powers of $\t-1$.
In this manner we arrive at the following approximation of the
precursor dynamic behavior at its front:
\begin{equation}\label{e33}
\widetilde{E}^S(z,t)\approx \frac{4\: 2^{1/4}\b\o_c
(\t-1)^{3/4}(b-6\d\sqrt{2(\t-1)})[1-2\d\l(\t-1)]}
{b^{7/2}\sqrt{\pi\l}}.
\end{equation}
In Fig.\ 4 an example of the precursor dynamics for $\b=1.0\times
10^{16} s^{-1}$, its envelope as given by (\ref{e32}) and the
envelope approximation as given by (\ref{e33}) are shown.

\begin{figure}
   \centering
      \includegraphics[width=.7\textwidth]{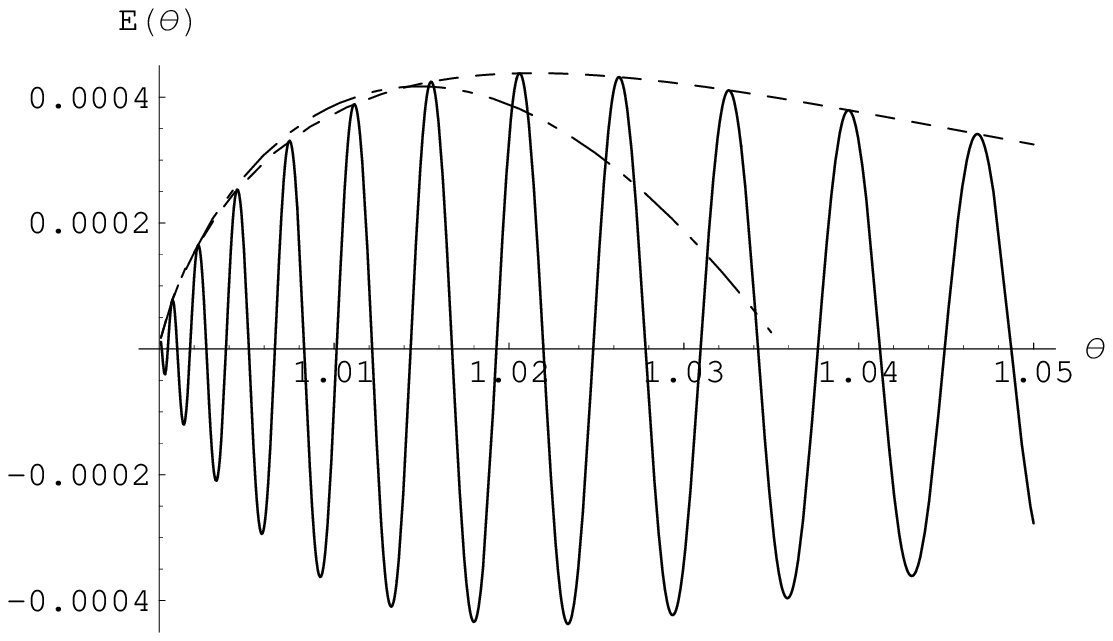}
      \caption{\textit{Sommerfeld precursor, its
       envelope and the envelope approximation at the precursor front.
       Calculated for $\b=1.0\times 10^{16} s^{-1}$, $\l=5.0\times
       10^{-15} s$ and $\o_c=2.0\times 10^{16} s^{-1}$.}}
\end{figure}

\vspace{1.5ex} The slope of the envelope is given by
\begin{equation}\label{e34}
\frac{d\widetilde{E}^S(z,t)}{d\t}\approx
\frac{2^{1/4}\b\o_c\{b\;[3-14\l\d(\t-1)]+
6\d\sqrt{2(\t-1)}[18\l\d(\t-1)-5]\}}
{b^{7/2}\sqrt{\pi\l}\;(\t-1)^{1/4}}.
\end{equation}
For sufficiently small $\t$, the terms proportional to $\d$ can be
neglected to yield
\begin{equation}\label{e35}
\frac{d\widetilde{E}^S(z,t)}{d\t}\approx
\frac{3\:2^{1/4}\b\o_c}{b^{5/2}\sqrt{\pi\l}\;(\t-1)^{1/4}}.
\end{equation}
It is seen that the slope of the precursor envelope steadily
decreases with growing $\t$.

\textsf{The case of large $\b$}

\vspace{1ex} \noindent For finite $\o^+$ and sufficiently large
$\b$, the arguments in the functions $\bet$ in (\ref{e3a}) can be
made arbitrarily small. Then the asymptotic expansion is
\begin{equation}\label{e36}
\bet(u)=\frac{1}{u}+\ln{2}+\frac{\pi^2 u}{12}+O(u^2), \qquad
u\rightarrow 0.
\end{equation}
With its use in (\ref{e25}), the approximation for
$G(\o^+;\b,\o_c)$, appropriate for the third region in Fig.\ 3,
follows:
\begin{equation}\label{e37}
G(\o^+;\b,\o_c)\sim \sqrt{\frac{\hbox{Re}(\Psi(\o^+,\t))}
{\Psi_{\o\o}(\o^+,\t)}}\cdot
\frac{-\o_c\;[24\b^2+\pi^2(\o^{+^2}-\o_c^2)]}
{12\b^2(\o^{+^2}-\o_c^2)}.
\end{equation}

\noindent Proceeding as in the previous case, we find the
following approximation for the envelope of the Sommerfeld
precursor
\begin{equation}\label{e38}
\widetilde{E}^S(z,t)\approx
\frac{[1-2\d\l(\t-1)]\;[b^2\pi^2(7+\t)+4(\t-1)\;
(96\b^2+\pi^2\o_0^2)]\;\o_c}
{192\;2^{1/4}\;b^{3/2}\;\sqrt{\pi\l}\;\b^2\;(\t-1)^{3/4}},
\end{equation}
where minimal $\t$ (here, $\t\approx 1.00003$) corresponds
to the first extreme of $J_1(x)$, occurring at $x\approx
1.841$.

Fig.\ 5 shows the precursor dynamic behavior for $\b=1.0\times
10^{19} s^{-1}$, its envelope and the approximation to the
envelope as given by (\ref{e38}). Notice that for any finite $\b$
there exists a $\t$, below which the assumption of large $\b$ and
moderate $\o^+$ is no longer valid. Therefore one should expect
that with rising $\t$, the precursor dynamics may pass through the
stages described by the first or/and second regions in Fig.\ 3,
before it reaches the stage characteristic of the third region.

\begin{figure}
   \centering
      \includegraphics[width=.7\textwidth]{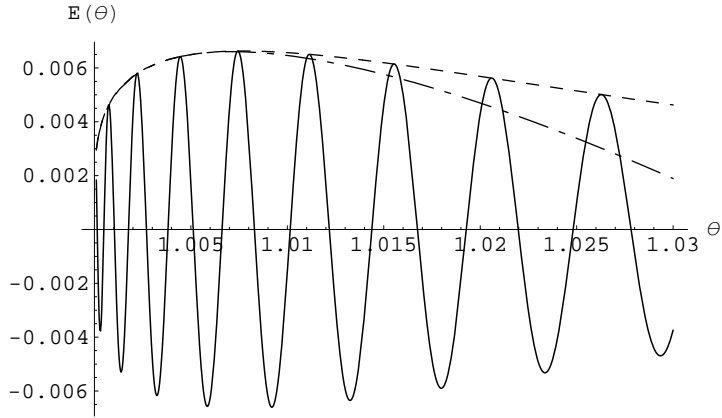}
      \caption{\textit{Sommerfeld precursor, its
        envelope and the envelope approximation at the precursor front.
        Calculated for $\b=1.0\times 10^{19} s^{-1}$, $\l=5.0\times
        10^{-15} s$ and $\o_c=2.0\times 10^{16} s^{-1}$.}}
\end{figure}

\vspace{1.5ex}  The slope of the precursor front is
\begin{eqnarray}\label{e39}
\lefteqn{\frac{d\widetilde{E}^S(z,t)}{d\t}\approx
\frac{b^2\pi^2\o_c[\t-25+2\d\l(3+2\t-5\t^2)]}
{768\;2^{1/4}b^{3/2}\sqrt{\pi\l}\;\b^2(\t-1)^{7/4}} } \\
\nonumber   & &
{\hspace{1in}}+\frac{\o_c(\t-1)[1-10\d\l(\t-1)](96\b^2+\pi^2\o_0^2)}
{192\;2^{1/4}b^{3/2}\sqrt{\pi\l}\;\b^2(\t-1)^{7/4}}.
\end{eqnarray}
If terms proportional to the parameter $\d$ are neglected,
this expression reduces to
\begin{equation}\label{e40}
\frac{d\widetilde{E}^S(z,t)}{d\t}\approx
\frac{\o_c[b^2\pi^2(\t-25)+4(\t-1)(96\b^2+\pi^2\o_0^2)]}
{768\;2^{1/4}b^{3/2}\sqrt{\pi\l}\;\b^2(\t-1)^{7/4}}.
\end{equation}
As before, the envelope slope decreases with $\t$. For the
parameters used here the rate of the precursor growth is about 16
times higher than in the previous case.

If we formally let $\b$ tend to infinity, we obtain from
(\ref{e39})
\begin{equation}\label{e41}
\frac{d\widetilde{E}^S(z,t)}{d\t}\approx
\frac{2^{3/4}\o_c[1-10\sqrt{\d\l}\;(\t-1)]}
{b^{3/2}\sqrt{\pi\l}\;(\t-1)^{3/4}},
\end{equation}
which corresponds to the unit-step function envelope in
the initial signal.

Finally, let us consider the precursor behavior at the first
oscillation, provided $\b$ is finite. With the use of (\ref{e28}),
(\ref{e31}) and power expansions of the Bessel and exponential
functions, we find for $\t\simeq 1$ (see Fig.~6)
\begin{equation}\label{e42}
\widetilde{E}^S(z,t)\approx \b\l\o_c(\t-1)^2
\left(\l-\frac{24\;\d}{b^2}\right).
\end{equation}

\begin{figure}
   \centering
     \includegraphics[width=.7\textwidth]{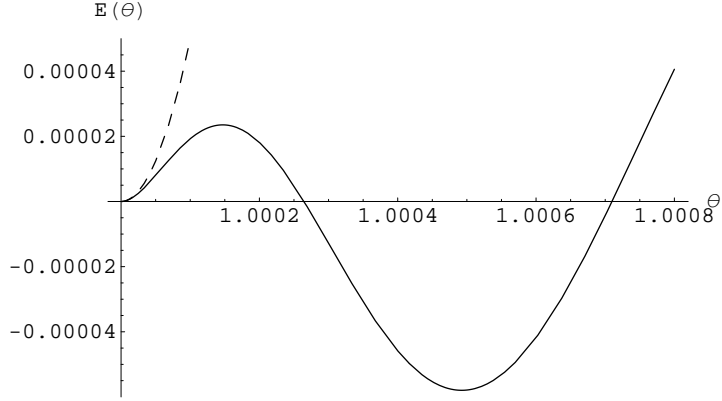}
     \caption{\textit{Onset of the Sommerfeld precursor calculated
     for $\b=1.0\times 10^{16} s^{-1}$, $\l=5.0\times 10^{-15} s$
     and $\o_c=2.0\times 10^{16} s^{-1}$. The dashed line
     approximates the precursor growth near $\t=1^+$ (Eq.
     \ref{e42}).}}
\end{figure}

\noindent This implies that
 \[\displaystyle \lim_{\t \to 1^+} E^S(z,t)=\lim_{\t
\to 1^+}\frac{\partial E^S(z,t)}{\partial t}=0, \qquad \lim_{\t
\to 1^+}\frac{\partial^2 E^S(z,t)}{\partial t^2}\ne 0. \]

\noindent On the other hand, we obtain
\[\displaystyle \lim_{t \to
0} E_0(t)=\lim_{t \to 0}\frac{d E_0(t)}{d t}=0, \qquad \lim_{t
\to 0}\frac{d^2 E_0 (t)}{d t^2}\ne 0. \]

\noindent It then follows that the Sommerfeld precursor has the
same smoothness properties about $\t=1$ as the initial signal
(\ref{e2}) has about $t=0$. This confirms the more general results
based on Green's function approach \cite{ri;97} and \cite{hs;98}.

\subsubsection{Special case}

Assume that $\b\rightarrow\infty$ and $\t\simeq 1^+$. Then
$g(\o^+;\b,\o_c)\approx -2\o_c\sqrt{2(\t-1)}b^{-2}$, and (\ref{e24})
reduces to
\[ E^S(z,t)\approx \frac{\o_c\sqrt{2(\t-1)}}{b}J_1(\l
b\sqrt{2(\t-1)} ). \] It is readily seen that this result fully
agrees with the representation
\[ E^S(z,t)\approx
\frac{2\pi}{\tau}\;\sqrt{\frac{\mathfrak{t}}{\xi}}\;
J_1(2\sqrt{\mathfrak{t}\xi}) \] obtained by Sommerfeld on the
grounds of integral considerations and valid for the initial
signal described by the Heaviside unit step function
(\cite{br;60}, Eq.~(33)). Here, we have employed Sommerfeld's
notation:
\[\displaystyle \mathfrak{t}=t-\frac{z}{c} \qquad
\xi=\frac{b^2 z}{2c} \qquad \tau=\frac{2\pi}{\o_c}.\]

\subsubsection{A comment on the form of asymptotic representation
of the precursor}

The form of the asymptotic representation of the Sommerfeld
precursor depends on the way $E_0(t)$ behaves at $t\approx 0^+$.
First we note that, (\cite{ja;75}, Eqs.\ (7.128)), the asymptotic
behavior of $E_0(t)$ just after it is turned on:
\begin{equation}
E_0(t)\sim \frac{a t^r}{r!} \quad \hbox{as} \quad t\rightarrow 0^+
\end{equation}
implies the following asymptotic behavior of $g$ at infinity:
\begin{equation}\label{e45a}
g(\o\;;\b,\o_c)\sim \frac{a}{2\pi} \left(\frac{i}{\o}\right)^{r+1}
\quad \hbox{as} \quad |\o|\rightarrow \infty.
\end{equation}
If the RHS of (\ref{e45a}) is used in the basic integral formula
describing the signal evolution in a dispersive medium, it appears
(\cite{ja;75}, Eqs.\ (7.144)) that in the vicinity of the front,
i.e.\ for $\t\rightarrow 1^+$, the Sommerfeld precursor dynamics
is described by $J_r[b\sqrt{2(\t - 1)}z/c]$. If
$\b\rightarrow\infty$, one has $r=1$ (which follows from expanding
the sine function alone), and the precursor dynamics is described
in terms of $J_1$. If $\b$ is finite, $r=2$ (see below
(\ref{e42})), and now $J_2$ describes the precursor development.
The orders of the Bessel functions appearing in the uniform
asymptotic representation of the precursor are determined from a
similar criterion, relating these orders to the behavior of the
integrand at infinity. In our case the smallest value of $r$ is 1,
which implies the presence of the functions $J_1$ and $J_2$ in the
asymptotic formula (\ref{e24}).

With increasing $\t$, the real value of $\o^+$ in the arguments of
beta functions in (\ref{e3a}) decreases. This results in moving
the boundaries between the three characteristic regions of
$G(\o^+;\b,\o_c)$ variation with $\b$ in the direction of smaller
values of $\b$. As a consequence, $J_1$ describes the precursor
behavior not only at $\b\rightarrow\infty$, but also at finite,
sufficiently large values of this parameter (see Fig.~3).

\subsection{Decay of the precursor}

By expanding the phase function $\Psi$ in terms of powers
of $\d$ we obtain
\begin{equation}\label{e43}
\Psi(\o^+,\t)=\o^+
\left(\sqrt{1-\frac{b^2}{\o^{+^2}-\o_0^2}}-\t \right)+
\frac{i b^2\o^{+^2}\d}{(\o^{+^2}-\o_0^2)^2
\sqrt{1-\frac{b^2}{\o^{+^2}-\o_0^2}}}+ O(\d^2).
\end{equation}
Since the imaginary part of the complex frequency $\o^+$ is small
compared to the real part of this frequency, the imaginary part of
the phase function $\Psi$ can be approximated with the second term
in this expansion.

For $\t$ small, $|\o^+|>>\d$, and we neglect $\d$ in the saddle
point equation (\ref{e7}) to arrive at
\begin{equation}\label{e44}
\frac{b^2\o^{+^2}}{(\o^{+^2}-\o_0^2)^2}+
\frac{\o^{+^2}-\o_1^2}{\o^{+^2}-\o_0^2}=\t
\sqrt{\frac{\o^{+^2}-\o_1^2}{\o^{+^2}-\o_0^2}}.
\end{equation}
For $\t$ not too large, the square root at the RHS in this
equation, and consequently the second term at its LHS, as well as
the square roots in (\ref{e43}), can be approximated by 1. It
follows then that the first term in (\ref{e44}), appearing also as
a factor in the second term of (\ref{e43}), equals approximately
$\t-1$. Thus by (\ref{e43}),
\begin{equation}\label{e45}
\hbox{Im}\;\Psi(\o^+,\t)\approx \d (\t-1).
\end{equation}
The function $\hbox{Im}[\Psi(\o^+,\t)]$, and its
approximations given in (\ref{e43}) and (\ref{e45}) are
shown in Fig.\ 7.

\begin{figure}
   \centering
     \includegraphics[width=.7\textwidth]{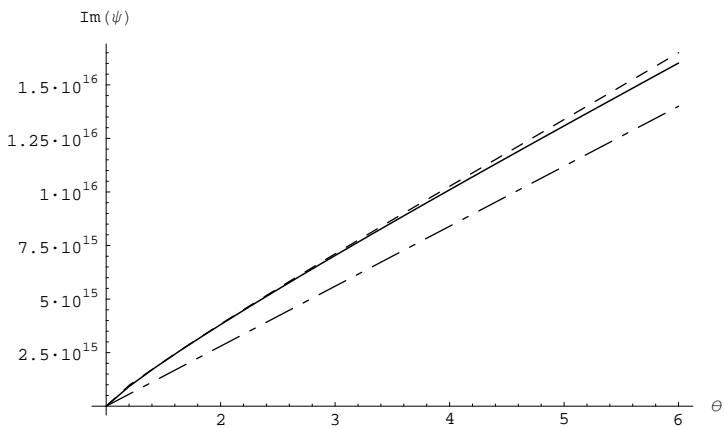}
     \caption{\textit{Comparison of
      $\hbox{Im}\;\Psi(\o^+,\t)$ (solid line) with the approximations
      given in (Eq.\ref{e43}) (dash-dash line) and (Eq.\ref{e45}) (dash-dot
      line), respectively.}}
\end{figure}

\vspace{1.5ex} It follows that major cause of the precursor decay
is its exponential damping. For low and moderate $\t$, the
exponent determining the rate of decay is approximately equal to
$-\l\d(\t-1)$. The contribution to precursor decrease due to
natural lowering of the oscillation amplitudes in the Bessel
functions is of smaller importance.

\section{The location of the saddle point $\o^+(\t)$}\label{s}

As seen from (\ref{e24}), evolution of the precursor depends on
the function $\o^+(\t)$ which is a solution to SPE, describing the
location of the distant saddle point of (\ref{e4}) in the right
complex $\o$ half-plane. Since this function does not seem to be
expressible in a closed form, approximate formulas for it were
found in the literature (Brillouin \cite{br;60}, Oughstun and
Sherman \cite{ou;97}, and Kelbert and Sazonov \cite{ks;96}). In
this section we offer another approximate solution to (\ref{e7}),
which is more accurate than the previous ones.

We begin with approximate solution to SPE found in \cite{ac;99}
which provides good accuracy for large and medium $\t$, but fails
for $\t$ close to 1. First let us notice that $\o^+(\t)$ can be
written down in the form
\begin{equation}\label{e8}
   \o^+(\t)=\sqrt{\o_0^2-\d^2+\z^{-1}(\t)}-i\d,
\end{equation}
where
\begin{equation}\label{e9}
 \z(\t)=\frac{1-n^2(\t)}{b^2}.
\end{equation}
The approximation obtained in \cite{ac;99} results from
substituting for $n(\t)$
\begin{equation}\label{e14}
n\approx g-\sqrt{\frac{1}{4 a}\left(\frac{\t}{g}-2 c\right)-g^2},
\end{equation}
which leads to approximation of $\z(\t)$, to be denoted by $\z_2$.
In (\ref{e14}),
\begin{equation}\label{e15}
g=\frac{1}{2\sqrt{3
a}}\sqrt{\frac{2^{-1/3}}{(u+\sqrt{u^2-v^3})^{1/3}}
\left[(u+\sqrt{u^2-v^3})^{2/3}+v\right]-2 c},
\end{equation}
\begin{equation}\label{e16}
u=2 c^3-72 a c e + 27 a \t^2,
\end{equation}
and
\begin{equation}\label{e17}
v=2^{2/3}(c^2-12 a e).
\end{equation}
The coefficients $a$, $c$ and $e$ are constant for a given medium
and are equal to
\begin{equation}\label{e18}
a= \left[\o_0^2-\d^2+\frac{i\d b^2}{\sqrt{\o_1^2-\d^2}}-
i\d\sqrt{\o_1^2-\d^2}- \frac{i\d b^2 [3
b^2+4(\o_0^2-\d^2)]}{8(\o_1^2-\d^2)^{3/2}}\right] \frac{1}{b^2},
\end{equation}
\begin{equation}\label{e19}
c=-\frac{i\d}{2\sqrt{\o_1^2-\d^2}}-
\frac{2}{b^2}\left(\o_0^2-\d^2-i\d\sqrt{\o_1^2-\d^2}\right),
\end{equation}
and
\begin{equation}\label{e20}
e=\frac{\o_1^2-\d^2-i\d\sqrt{\o_1^2-\d^2}}{b^2}.
\end{equation}
We shall denote this approximation by $\o^+(\t,\z_2)$.

For $\t\approx 1^+$, i.e.\ in the vicinity of the precursor front,
we find here another approximation. By expressing SPE in terms of
the variable $\z$ we obtain
\begin{equation}\label{e10}
  w \z-i\d\sqrt{\z}\sqrt{1+w \z}=\frac{\t n-1}{b^2 \z} \;,
\end{equation}
where $w=\o_0^2-\d^2$. If $\t\rightarrow 1^+$ then $\z\rightarrow
0$ and this equation can be approximated by
\begin{equation}\label{e11}
   w \z-i\d\sqrt{\z}=\frac{\t-1}{b^2 \z}-\frac{\t}{2} \;.
\end{equation}
The solution to (\ref{e11}), relevant to our problem can be found
by means of the \textit{Mathematica} computer program:
\begin{equation}\label{e12}
  \z_1=\frac{1}{2w}\left(u-\frac{q}{2}+\sqrt{2u^2-3s-
  \frac{q^3-\frac{j}{b^2}q-\frac{8h}{b^2q}}{4u}}\right),
\end{equation}
where {\setlength\arraycolsep{2pt}
\begin{eqnarray}\label{e13}
  q&=&\frac{v}{w}, \quad j=b^2\t^2-8(\t-1)w, \quad
  h=-l\t(\t-1), \\ l&=&\d^2+\t w, \quad u=\sqrt{s+\frac{q^2}{4}-
  \frac{j}{6b^2}}, \quad  s=\frac{1}{6b^2}\left(\frac{z}{2^{2/3}p}+
  \frac{p}{2^{4/3}}\right),   \nonumber \\[1ex]
  p&=&(r+\sqrt{r^2-4z^3})^{1/3}, \quad
  z=b^4[j^2-4\;12b^2h+4^2\;12w^2(\t-1)^2],  \nonumber \\[1.5ex]
  r&=& b^6[2j^3-12^2b^2jh-12^2(\t-1)^2(8jw^2-12b^2l^2-12b^2w^2\t^2)].
  \nonumber
\end{eqnarray}}
By substituting $\z$ for $\z_1$ in (\ref{e8}), a new approximation
of the distant saddle points location is obtained which is valid
for $\t$ close to unity. We denote it by $\o^+(\t,\z_1)$.

The two approximations, $\o^+(\t,\z_1)$ and $\o^+(\t,\z_2)$, can
now be combined into one formula that provides smooth transition
from one approximation to the other. For Brillouin's choice of the
medium parameters we choose the transition value of $\t$ to be
$1.3$. Then the joint approximation can be written down as
\begin{eqnarray}\label{e21}
\lefteqn{\o_{SD}(\t)=\o^+(\t;\z_1)\left[
H(1.3-\t)+\frac{\mbox{sign}(\t-1.3)}{2}
\frac{\kappa(\t-1.3)}{\kappa(0)}\right]} \hspace{1in} \\ & &
{}+\o^+(\t;\z_2)\left[H(\t-1.3)- \frac{\mbox{sign}(\t-1.3)}{2}
\frac{\kappa(\t-1.3)}{\kappa(0)}\right],  \nonumber
\end{eqnarray}
where $H(s)$ is a unit-step function,
\begin{equation}\label{e22}
\kappa(s)=\eta(s+0.05) \eta(s-0.05),
\end{equation}
and
\begin{equation}\label{e23}
\eta(s)=\left\{
\begin{array}{ll}
\exp{(-1/s^2)}, & s>0, \\ 0, & s\le 0.
\end{array}
\right.
\end{equation}
Since $\kappa(\t-1.3)$ is zero outside the interval
$1.25<\t<1.35$, the approximation $\o_{SD}(\t)$ equals
$\o^+(\t,\z_1)$ if $\t<1.25$, and $\o^+(\t,\z_2)$ if $\t>1.35$.
The factors in square brackets in (\ref{e21}) provide smooth
transition between the two approximations inside the interval. At
$\t=1.3$, they are understood in a sense of limits (from the left
or from the right), and thus are equal to $1/2$. A different
choice of media parameters may require the numerical parameters in
(\ref{e21}) and (\ref{e22}) to be modified.

\begin{figure}[ht]
   \centering
     \includegraphics[width=.7\textwidth]{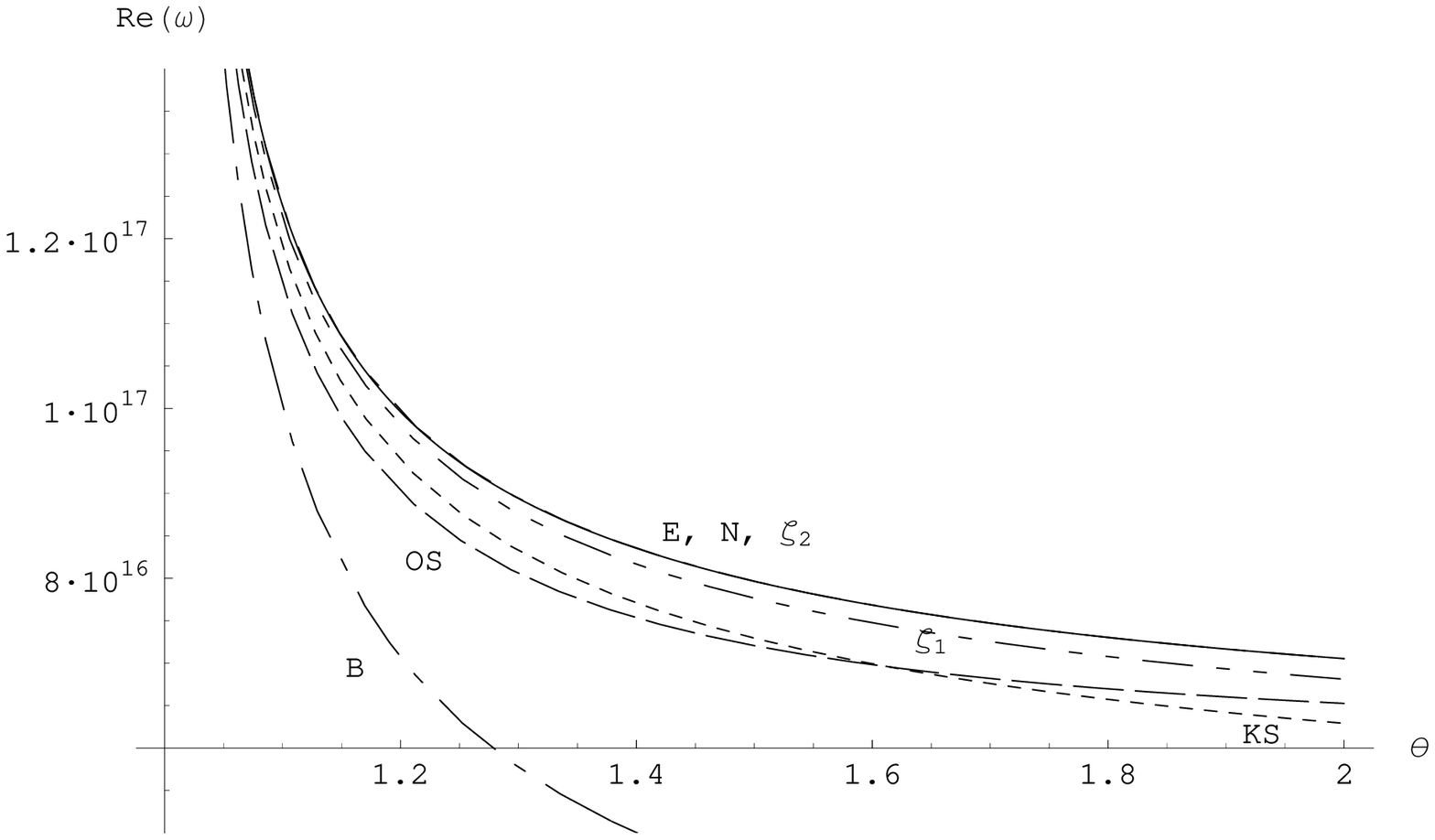}
     \includegraphics[width=.7\textwidth]{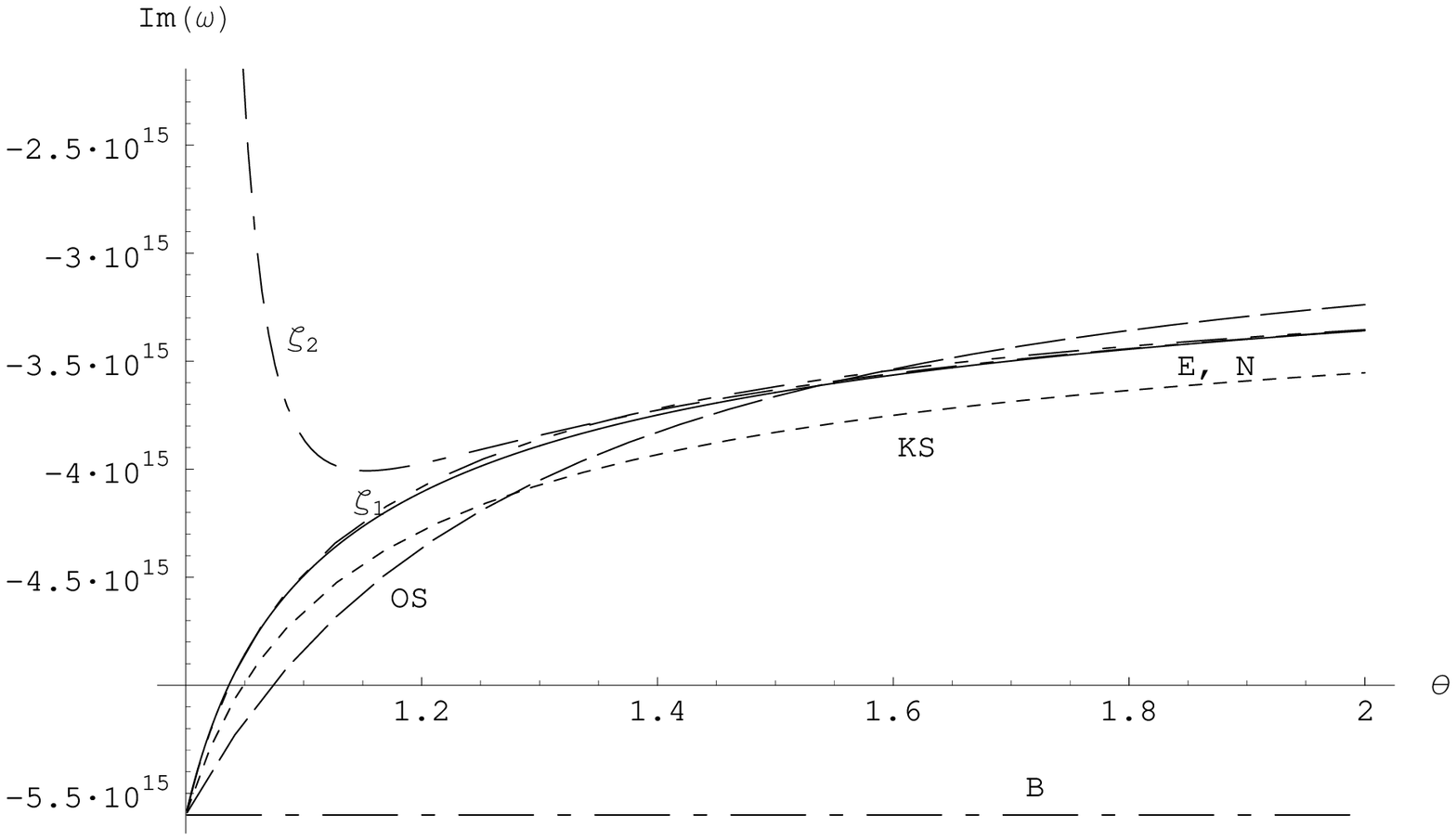}
     \caption{\textit{Real and imaginary parts of various
     approximate solutions to (Eq.\ref{e7}). Legend: E -- exact
     result (solid line), B -- Brillouin's approximation, OS -
     Ougstun-Sherman's approximation, KS -- Kelbert-Sazonov's
     approximation, N -- the result based on (Eq.\ref{e21}),
     $\xi_1$ and $\xi_2$ -- the approximations described by
     $\o^+(\t,\xi_1)$ and $\o^+(\t,\xi_2)$, respectively.
     }}
\end{figure}

\vspace{2ex} The approximation (\ref{e21}) is shown in Fig.8,
together with the approximations obtained by Brillouin, Kelbert
and Sazonov, Oughstun and Sherman, and with the partial
approximations $\o^+(\t; \z_1)$ and $\o^+(\t; \z_2)$. This
approximation is not as simple as the first three ones, but may
prove to be useful when a higher accuracy is required. Its maximal
deviation from the solution of (\ref{e7}) found numerically
slightly exceeds $2$ percent.

Note that the approximation $\o^+(\t; \z_2)$ indicates an
interesting symmetry between locations of the distant and near
saddle points (see \cite{ac;99}). Similar symmetry also follows
from the Kelbert-Sazonov approximation.

\vspace{1.5ex} \noindent \textbf{Acknowledgment}

\noindent The research presented in this work was partially
supported by the State Committee for Scientific Research under
grant 8 T11D 020 18.


\end{document}